# Energy-Length Rule.

[+]Alexandru C Mihul, Eleonora A Mihul
Bucharest University

**Abstract.**

Lorentz ordering (causality) implies the following rule: for any given energy $p_0$ of a system there is a certain interval $\chi^0$ on $x^0$ so that their product is the Lorentz ordering constant L It means $p_0\chi^0 = L$. The constant L=hc. Hence Planck constant h in a similar way as c are both consequences of Lorentz metric. The basic ideas are:

1. Lorentz metric implies that $x^0$ must represent a length like the other components of x in X

2. The dual metric space X* is well defined since the Lorentz metric tensor is not singular. The components of the vectors p in X*are interpreted as representing energy. The properties of the physical systems that are direct consequences of the detailed structure of X and X*, and so expressed through the Lorentz Limit L are presented.

[+]Corresponding address Alexandru.Mihul@cern.ch

# Energy-Length Rule.

## Lorentz metric and time concept.

(The mistake of interpreting $x^0$ as representing the concept of time)

> We, as sentient beings, never lose
> The consciousness of time.
> It seems that the nature laws
> Do not care about our consciousness.
> Then what about the laws of Reality?.

**Introduction.**

The purpose of the paper is to find out how much of the competent critiques of our theories, so clear presented by Roger Penrose in his book "The Road to Reality "were avoided if the misleading concept of time would have not been introduced in special relativity, about a century years ago. It is referring only to the special relativity, (constant c) and quantum mechanic (constant h).
.

Let us first cite some texts from the book. Roger Penrose writs:
1." In my opinion, the theory of special relativity was not yet complete, despite the wonderful physical insights of Einstein and the profound contributions of Lorentz and Poincarė, until Minkowski provided his fundamental and revolutionary viewpoint: *spacetime*"{pg.406}.

2. "Accordingly the 'time' that a photon experiences (if a photon could actually have experiences) has to be zero!" (pg.407)

3. "An extreme situation arises when we have what is referred to as *causality violation* in which 'closed time like curves' can occur, and it becomes possible for a signal to be sent from some event into the past of that same event!" (pg.409}.

4. "It seems to me that the need for such coherence, in any proposed physical model, is unarguable." (pg.1014).

5. "I have deliberately refrained from addressing, at any great length, the question of conscious mentality in this book, despite the fact that this issue must ultimately be an important one in our quest for an understanding of physical reality." (pg.1030}.

6. "If the 'road to reality' eventually reaches its goal, then in my view there would have to be a profoundly deep underlying simplicity…". (pg.1034).

**The replica.**

Indeed, simplicity, inductive logic and coherency are underlying our approach which proves:

**First**:
Lorentz ordering (causality) implies the following rule: for any given energy $p_0$ of a system there is an interval $\chi^0$ on $x^0$ so that their product is the Lorentz ordering constant L It means $p_0\chi^0 = L$. Confronting with empirical facts one finds that $\chi^0$ is just what we call Compton wave length for massive systems and so $L = hc$, the product of Planck constant h and the "speed" of light c. The dimension of L is energy multiplied by length.
We show that the properties of photon and neutrino, as well as Newton and Coulomb laws, are consequences of the above rule.
The very concept of commutativity for photon and anticommutativity for neutrino, multiplicative factor of the unity being L, is provided by the ordering (causaly) defined on the matrix representation of the Clifford $C_2$ algebra.
Conceptually, all is relying on the fact that time has not a physically meaningful compatible with a space structured by Lorentz metric. It is an outside concept which serves to define the constant c as having the dimension of distance /time.

**Second** (as the consequence of the First)
Lorentz ordering (causality) implies a constant $L=hc$. It means that Planck constant h in a similar way as c are both consequences of Lorentz metric.
The basis idea is that faithful to the Lorentz interval which defines the length between two points, x' and x'' in a real 4-dimensional linear space X, through an additive relation of their components (squared), a coherent interpretation is that the all four terms (components of x'-x'') have also to represent, length.
The formal way in which Minkowski introduced time just by relabeled $x^0$, one of the four components of $x \in X$ by ct, and declaring that c having dimension of speed, t represents time, violates conceptually the very physical content of the Lorentz interval.
Hence the coherent starting point is to consider that all components of the four dimensional vector $x \in X$ are lengths.
Now, the conscious mentality is unavoidable issue in our understanding of physical reality. Inductive logic endowed us with the "ordering tool" with which we quest for understanding of physical reality. This is that makes us to assert that only the ordered points (what we are accustomed to call causally related events) in X are physically meaningful.
Therefore, since Lorentz causal logic manifest actually the inductive logic of the physicist as a sentient being or how Eddington coined her or him, the knower, makes Lorentz ordering (causality) to be inevitable , and so must be its implications.

### Lorentz Ordering (Causality).

Mathematically, Lorentz interval (metric) is just a particular representation of the mappings, providing partially ordered structured spaces (appendix) [1].

We denote by X a real linear space endowed with Lorentz metric.

Two points $x' = (x'^0, x'^1, x'^2, x'^3,) = (x'^0, \mathbf{x'})$ and $x'' = (x''^0, \mathbf{x''})$ are ordered if and only if the interval

$$\text{Int.}(x'-x'') = (x'^0 - x''^0)^2 - (\mathbf{x'} - \mathbf{x''})^2 > 0$$

either

$$\text{Int.}(x'-x'') = (x'^0 - x''^0)^2 - (\mathbf{x'} - \mathbf{x''})^2 = 0$$

and

$$x'^0 > x''^0$$

is satisfied.

The space X is a partially ordered space defined by existence of two quite equivalent cones; up with $x^0 > 0$, and down with $x^0 < 0$.

Between the curves that relate two ordered points in the space X, there is a longest one. It means that for any two ordered points there is a continuum curve of ordered points, a longest one, extremal. Precisely, it is a Nonextendable Ordered (causal) curve (NOC). Each such a curve is actually a straight line parameterized by $\mathbf{a}/a^o < 1$, heaving as a limit $\mathbf{a}/a^o = 1$, $a^o \neq 0$, where $a(a^o, \mathbf{a})$ is a four dimensional vector in X. For the norm $\mathbf{a}/a^0 > 0$, the straight line is in the interior of the cone, and for the norm of $\mathbf{a}/a^0 = 0$ it provides what we call a ray which is on the boundary of the cone.

The limit $\mathbf{a}/a^o = 1$ stands for the existence of c that implies the time as an outside parameter.

Nowadays we represent physical system in the above structured space X. The very concept of "physical system" implies, at least, to bring besides in, the concept of energy.

Since the Lorentz metric is not singular the bilinear form

$$p(x'-x'') = p_0 x^0 - \mathbf{p} \cdot \mathbf{x} \qquad \qquad 1)$$

provides the opportunity to define the energy of a system represented in X.
(for brevity of writing we denoted x'-x'' by x). The $\mathbf{p} \cdot \mathbf{x}$ is the scalar product between 3-dimensional $\mathbf{p}$ and 3-dimensiona $\mathbf{x}$ (bilinear form of the components of $\mathbf{p}$ and $\mathbf{x}$) The bilinear form 1) defines what one calls metric dual space of X which we denote by X*.

The components of the vectors $p = (p_0, \mathbf{p})$, obviously 4-dimensional, are the coefficient of the functional 1).

The components of a vector p in the dual space, $p \in X^*$ we interpret as representing the energy of the system.

Obviously, the metric in X* is induced by the metric of X, as well as the dual group of the causal group of X.

The p(x) is invariant means: If there is a symmetry group G of transformations applied to x, which leaves invariant the metric of X [2], then

there is the dual group G* whose transformations applied to p, leave invariant
p(x) i.e. p(x)= G*p(gx).

For that we assert actually the existence of the bijection of physical systems and the space X.. It means that we assert meaningful to attach to real systems their representations in points of X. Hence any mathematical findings provided by a given structure of X and X* must have physical interpretation. This is what we are looking for.[3].

We call free systems., those representing by a certain p, whose invariant $p^2$, to G*, is positive or zero, and so, by p(x'-x''), for ordered x' and x ''belonging to NOC. We have to prove that such a structure exists.

Hence p(x'-x') defined above 1) expresses actually the principle of inertia:": every body has a weight in the direction of its movement " defined by Francis Bacon .

Before coming to the consequences of the above assertions let to make clear our position.

We don't advocate the ordering concept. We just look for its consequences without making any other assumption. Lorentz interval divides the space X in causally related points, - i.e. ordered - on the both light cones and on the boundaries of the cones, and in not ordering points, out of the cone (space-like interval).

It is important for what follows that Lorentz ordering relation is an ordering relation in X with regard to which X is directed [4] It means that for any x' and x'' in X, there is an x''' in space X so that x''' is ordered related to x' and to x''. If X is an ordered linear space, then X is directed if and only if the cone generates X, the Clifford theorem [5]. It reflects the exquisite, profound role of the light.

Certainly, the structure of X by itself cannot provide properties of a physical system.

However, significant clue to the problem is that the Lorentz metric tensor is not singular, (for Newtonian affine spacetime it is singular). So its dual metric space is precisely defined. The components of a four-dimensional vector p of the dual space X* of X, are the coefficients of the real-valued linear functional 1).

Actually the relation 1) is a particular example of Weyl proposes of the locality.

Hence, we are factually dealing with two dually "related" spaces X and its dual, X*.

Now we prove that the invariance of p(x) under the translation group, subgroup of the ordered Group [5], implies a structure describing free physical systems .

**The prove**.

"Energy momentum conservation"[3].

Now we associate to a particle a 4-vector p from X*and an oriented (causal) line NOC in X.

The point is: can we associate to the same NCC two vectors p and q in X ???
Let us assume Yes and prove no.

For that let us take any given two points on a NOC, x and y, then p(x) and q(y) (≡ $q_0 y^0$ – **q.y**_). are two real numbers**.** We can have any of two possibility for p(x)> q(x), or less. The result is independent. Let us take
$$p(x) < q(y).$$
Invariance to the translation group T (the subgroup of the causal group) means

$$p(x + a) < q(y + a) \text{ for any } a \in T$$

Being a linear functional we have

$$px + pa < qy + qa$$

that is

$$(p - q)a < qy - px$$

or explicitly,

$$(P_0 - q_0)a^0 - (\mathbf{p-q}).\mathbf{a} < \text{given number}$$

Since the right side is a given number, and the inequality must be valid for any translation $a \in T$, (the group T being a subgroup of the causal group) each component of p and q must be equal. What one can call the conservation of the 4-dimensional vector p.

    Let us now define the concept of the free system, it means:
    I. a certain 4-dimensional vector p,
    II. x belonging to a NOC in X (ray is a particular case);
    III. $p^2 = Y$, $p^2 = p_0^2 - |\mathbf{p}|^2$, the invariant Y to the dual causal group is positive or zero;
    IV. the invariant $Y = 0$ i.e. $p_0 = |p|$ if and only if x belongs to the ray i.e.
    $x_0 = \pm |x|$

**Definition**.: That system (particle) representable by such a structure, we call Free System (FS).

    Hence, since a real physical system is described simultaneously in the both spaces, and since the invariant p(x) is the only entity satisfying this condition, p(x) has to play an essential role in our investigation. The ordering (causality) constraint excludes zero value for p(x) (orthogonality condition) for Y positive or zero, for the isolated systems.

    In those cases for ordered x' and x'', p (x' – x '') is a sequence defined on a ordering subspace of X, precisely on NOC.

    According to E. H. Moore and H.L. Smith [6] such a sequence (filter) defined on an ordered set in a space directed with respect to the ordered relation is convergent and has a limit. This limit provides a "constant" which we call the Lorentz Constant (LC) and denote it by L.

    It means that for any given p with either $p^2 > 0$ or with $p^2 = 0$, there is a certain vector $\chi$ with the components $(\chi^0, \chi^1, \chi^2, \chi^3)$ for which

$$P_0 \chi^0 - p_1\chi^1 - p_2\chi^2 - p_3 \chi^3 = L \qquad 2)$$

    It is important to attach a physical meaning to L. For that, we have to choose measurable entities for the components of x and for the components of p.

    Surprise or it is trivial? Since the Lorentz interval is a sum of the components (squared) of x, it is obvious that all components must represent the same entity. We chose length for all four components of x and, for all components of p we choose energy.

    Then, the dimension of L is length multiplied by energy (just the dimension of the square of electric charge, also consequence [7] of the above structure as we shall show later.

In our mind's eye we find something that has meaning to us, but unless we have objective proof of checking it, we don't know if that something has any meaning. Therefore, the justification of Lorentz ordering, even if it is motivated as representing inductive logic, rests in the proof of physically meaning of the theoretical findings 2) [7].

Then we have to find those properties of the FS that are direct consequences of the detailed structure of X and X*, and so expressed through the Lorentz Limit. L

In particular for $p^2 > 0$ rest ("massive") particle, for p=0 compatible with the above constraints we find that $x^o$ represents a natural scale

$$\chi^0 = L / p_0 \qquad 3)$$

attached to any such a particle, . Confronting with empirical data, we find that we got the scale attached to a rest ("massive") particle, just what we call Compton wavelength $\lambda$ = h/mc for $p_0 = mc^2$ and L = hc, is a ordered (causal) constant.

The concept of mass cannot be defined in the above setting until we shall not define the constant c, whose definition implies the outer concept of time.

**Note:** As it is known, the wave description of matter defines a natural scale for a particle through its Compton wavelength. The relation 3) shows that for a space endowed with Lorentz metric, the constraints imposed by Lorentz causality imply such a scale without any other assumption.

**Consequence.** Planck constant h is related to Lorentz causality as much as c is. We use rest and restless instead of massive and massless, because having not defined time we cannot define c as a speed and neither mass.(We shall define time as a outside parameter and so the constant c).

It is outstanding that the above mathematical statements, structures, are valid either for $\chi^0 > 0$, or $\chi^0 < 0$ (the two cones) referring to + L and –L respectively . This provides the concept of system and anti-system, in particular, particle and antiparticle regarding to the relation to the space X.

**Remark. 1**: As we did not used any other assumption, the meaning or meaningless of the consequences can be seen as true or false of Lorentz causality, obviously within the sphere of its applicability. According to our understanding the limit of applicability is expressed in experimental information. Therefore it seems that the existence of the microwave background, and probably an equivalent neutrino low energy background of neutrino and the minimum rest energy of the electron are not compatible asymptotically with the rule ; the higher the energy the smaller the interval $\chi^0$ on $x^0$ so that their product is well defined. Even if such speculations are often used.

**Remark.2**. The usual interpretation of the Lorentz causality defined in the Minkowski space-time, is actually an ordered relation on time defined on the axis $x^o$ = ct of the real linear space structured by Lorentz metric. In Newtomian frame "causality" means: If the state of a physical system is given at a certain time, then its state at any other time is determined by the laws of nature. In both cases causality expresses actually an ordering relation relying on the concept of time represented on the real ordered axis.

The basic idea is that Lorentz metric defines the "ordering" independently of the concept of time.

**Restless systems.**

For any **x** there is an interval $x^0$ such that $|\mathbf{x}|/x^0 = 1$, and it implies that the dual vector has Y=0. The point at issue is to find the measurable properties of such free physical system manifestly expressed through the constant L.

Let us analyze the "massless" or better call them restless systems.

for which $p_0 = |\mathbf{p}|$ and $x^0 = \pm |\mathbf{x}|$. . In this case, clear that only the relative orientation between **p** and **x** is to be chosen. The exquisite tandem between these two "vectors" is the clue of the "dynamical" structure of the restless systems.

There are two invariant possibilities:

One is transverse, i.e. 3-dimensional **p** and **x** are orthogonal. This is what I call rest-less transverse FS. For this case, (the scalar product being zero in 2)),

$$p_0 \chi^0 = L \qquad \qquad 4)$$

which is just the blackbody Planck relation for energy. The handedness - 1 and +1 represent two signs of the two-orientation "helicity". A massless particle such as a photon; can spin only about the direction of the transport of its energy. The norm of 3-dimensional **p** is just the energy of the system. However, as a vector **p** has to be interpreted as angular momentum indicating the both possible rotations with respect to the ray. This provides both orientations of the closed paths and so it looks like the elements of the first homotopy group of U(1) space group. Using the fact that the finest topology of X, induces discrete topology on the ray [9], the winding number of homotopy group "measures" the distance "covered" by the photon. Infinitely connected topology allowed photon to be stable "forever". Topologically, is represented by a cylindrical space. Is this the root of U(1) symmetry of the photon?

We emphasize that the relation 3) and 4) are formally the same. The energy of the system multiplied by an interval $\chi^0$ on $x^o$ axis is equal to the constants L=ch. Thus $x^o$ got its own role in our special relativity.

Another rest-less system is "longitudinal", i.e.: **p** and **x**, for $x^0 > 0$ must be antiparallel, pointing in opposite direction, left handed, "helicity" equals –1/2 and so the neutrino properties, are obtained.

For $x^0 < 0$, they must be parallel with + 1/2, right-handed antineutrino. Thus a system, as free one, with properties of which we call neutrino and antineutrino follow directly from 2). It is obviously that the energy is positive for neutrino as it is for, so called, antineutrino.

Thus the change of the sign of energy for definition of particle and antiparticle is avoided.

The both restless systems are actually sliding spinning systems satisfying the energy-length rule.

Hence the Lorentz causal relation defines three classes of free isolated systems: a class of massive ones for which we have the relation:

$$p_0 \chi^0 = L \qquad \qquad 1)$$

and two classes of massless systems:
transversal for which we have the relation

$$p_0 \chi^0 = L \qquad \qquad 2)$$
formally equal to 1)
and longitudinal systems for which
$$p_0 \chi^0 = L/2 \qquad \qquad 3)$$
For massive system, $\chi^0$ [12] provides what we call Compton wavelength.
and for massless photon its wave length

Now let us discuss the subtle issue of dimensionality of the space X lodging simultaneously rest (massive systems), one species of transverse rest less and. three species of longitudinal rest less systems, photons and neutrinos. For each system is defined antisystem described by the down cone.

The transverse systems being defined by the outer product, the space X must be at least 4-dimensionality. Instead the exquisite role in describing longitudinal system, beside mass less is played by the inner product .**p.x**. This lowers the grade of a vector so can perfectly be defined in 2-dimensional space. So our space were 3-dimensional, if only longitudinal restless systems there would be. Since we have to use the same space for representing the systems, as it is implicated in the way we introduced the concept of the space as representing Lorentz ordering {Partially ordered) structure, longitudinal system that requires only the inner product must be represented in a 3-dimensional subspace of X .

Then the question is: how many linear independent 3-dimensional subspaces has X...The space X has seven 3- dimensional subspaces. One is space like, three are time-like subspaces and. three are singular subspaces [9]

Except time like subspaces can lodge longitudinal systems, with their properties being restless and longitudinal. Since there are three such subspaces there should be three diverse types of "neutrino" and obviously their antineutrino.

Some details [10]:

X is generated by a basis of one time-like and three space–like vectors... If we keep the time like (1,0,0,0) and choose one of the three possible combinations for instance (0,1,0.0) and (0,0,1,0) , we get the basis that generates one of three time-like 3-dimentional sub-spaces. In each of them, independently we can define one of the three restless longitudinal systems.

The 3-dimensional space-like subspace generated by three space-like basis vectors is excluded "by causality."

.Peculiar are three singular spaces whose basis includes one light-like vector for instance (1,1 0,0), which together with (0,0,1,0) and (0,0,0,1) generate a singular 3-dimensional singular subspace. Rather akin to the longitudinal system they could play a exquisite role in the neutrino mass problem.

Obviously there are three such singular 3-dimensional subspaces.

**Remark on the possible neutrino mass.**

One considers that "if neutrinos do have mass they can change the flavors". According to above findings the change of the flavors means an interaction that change the 3-dimensional subspace representing the respective neutrinos .If there is such an interaction and if on a solid experimental data will be proved that the neutrino possesses a property interpretable as $p^2 \neq 0$, then either it has not a well defined helicity or its "mass" is provided by representing the neutrino as the sum of a light vector and a singular vector of the akin subspace.

**Partner.** If we are looking now for a partner of neutrino, we have to find a system whose some properties are common.. Symptomatic for neutrinos is a certain sharp handedness , let say left for neutrino and right for antineutrino., and they are representable in a time-like 3-dimensional subspace  Obviously that the partner ,if exist must be a rest system otherwise would coincide with neutrino itself.

According to the above finding; we assert the same energy-length law for the free system representable by two points x' and x'', in the space X, in which two equal massive systems are carriers equal charge. We discuss the case for which the interval between points is light-like. We assume that he communication between "the points" can be represented by  energy $A_0$.Then we deal with a 4-vector A in the dual space of X .It is dual to the 4- vector between the two points x' and x'' in X space.

With 3-dimensional vector **A,** we construct two dimensional tensor. The antisymmetric tensor, curl **A,** represents electromagnetic tensor and symmetric tensor of **A** defined in X* provides symmetric tensor in X space what is called Ricci tensor. We note that the curl of the vector potential for negative charge, electron, is strictly left and it is the right handedness  for positron. That is why neutrino "chooses" electron as partner and antineutrino chooses the positron. Thereby electron must be "particle" like neutrino and positron "antiparticle" like antineutrino.

Note that alpha constant is a quotient between the scale $x^0$ when the charge is present in X and the scale $x^0$ for free particle carrying the charge system The interval  $(x'-x'')^0$ fitting  energy-length rule for  communication energy between the charges' localized in x' and x' respectively.

Square of this charge up to a constant alpha is equal to L.

**Remark** about commutator and anticommutator.:  We define now ordering ( causality ) on the 2x2 matrix representation of the SL (2,C) group, using just Pauli representation of the matrices  taking determinant of the matrices  instead of the interval of the respective vectors. We find that the matrices corresponding to **p** and **x** anticommute for longitudinal case and, commute for transverse case. The multiplying factor of unit matrix is just causal limit L according to 4).

. Bilinear character of the commutator and anticommutator is the expression of the bilinear character of the bilinear functional  p(x) involved by Lorentz ordering (causality).

It seems that not only Planck constant but also the very concept of "commutativity" follows from Lorentz ordering defined on the matrix representation of the Clifford algebra $C_2$. We have not to make use of the classical "canonical conjugate" concept. Moreover the multiplicative factor of unity matrix is L.

The fundamental sense of commutator and anticommutator physically and mathematically is got in taking into account that the scalar product graduates a vector with one dimension less and the vector product leads to the bivector.

So we succeeded in proving that Lorentz ordering, called usually Lorentz causality [1], implies a "causal " constant L. Using empirical data one gets L=hc with the dimension of energy multiplied by distance.

Clearly, we could not separately get the Planck constant h, which is energy multiplied by time, neither the "speed" of light which is length over time, since we have no time as far as we associated the length to the all four component of the four dimensional vector x∈X. It follows that if we choose an interval of time (whatever it is) for a given $x^0$ to get c, we must use the same interval of time, to get h.

Then, $\chi^0$ is just Compton scale (wave length) for systems whose invariant of 4-vector p in the dual space X*, of X, interpreted as energy, is positive i.e. $p^2 > 0$.

For $p^2 = 0$, we have $p_0\chi^0 = hc$ which is just the Planck law for "radiation", and for which $\chi^0$ is just what is called wave length of the photon of energy $p_0$.

For neutrino the product is ½ hc.

Hence $\chi^0$ is that length, scale, that multiplied by the energy $p_o$ of a free system equals to the "causal " constant L=hc is implied by the Lorentz metric.

Trying to understand how the constant c has to be defined coherent with Lorentz metric, we found that time plays its role only as an outside parameter providing the meaning of the constant c as length/ time dimension, representing the constraints IV in the definition of the free system imposed by Lorentz causality [1] .In annex 1 we tried to make the relation between two types of "times" in our conception

Now let us find: who is c?

We have been taught and we were teaching, that:

a) Relativity is an extension of classical physics to the realm where the velocity, v=c, must be regarded as finite.

Let consider two points A and B at N meters apart. And we measure n second interval of time in which a photon starting in A reaches the point B. The experiment tells us that independently of the energy of the photon the quotient N/n has the same value. This value we denote by c. We say that the photons are moving with the same "velocity" c. Nonsense! Since vis vivas expressed as ½ $mv^2$ defines the energy of a system moving with velocity v. And c does not participate to the photon energy. Hence, c is not velocity even so it has the dimensionality distance over time. The transport of the energy of light takes place according to the symptomatic "dynamical" structure of light .One kind of "massless perpetual mobile ",

Interpretation of c as velocity is according to the traditional measuring as space over time. Then either space or time, or both of them are misinterpreted .Probably James Joice would "nominate" c "speeding-less speeding".

b) Quantum mechanics is an extension to the realm where Planck's constant, h, is not zero .Learning from above we should rather say that the energy has to be used and not action. Indeed, a photon is a certain perpetual transport of a given energy which multiplied by a length is just hc.

Light is a sliding spinning system transporting its energy according to its own structure. It is spinning necessarily about its direction of propagation. It ignores, equally, time and distance. It seems that it is rather coherent to interpret that light provide an example of Mach's principle "the very notion of a background space should abandon". Certainly that the root of the ordering as a inductive logic is our concept of time. Obviously, this does not mean that we have to introduce time to get ordering.

**Conclusion.**
Once we accept the use of the real, linear space X endowed with Lorentz interval between two points x' and x'', Int.(x'-x'') = $(x'^0 - x''^0)^2 - (\mathbf{x'} - \mathbf{x''})^2$ , we
Have first to explore the luxury [8] structure offered by it without any other assumption.

We have to realize that it radically altered "human", Newtonian way of thinking and, with it, our way of describing those properties of physical system related to the space in which we represent them. This analysis led to the new resolutions that carried the seeds for new ways of thinking, although it could be done hundred years ago
Precisely, if we assert that it is meaningful to describe physical systems in causally related points of X then the reach structure of X has to provide all the properties of the systems related to this representation.

**Appendix**.
**Physical Systems and Causal Structure**

We give first general formulation of causality, on which we rely all our results.
So we consider the physical system set **S** and the relations between them **R** as a priori concepts, and postulate the following properties:
I. For any systems s∈**S**, there is the relation r∈**R** from s to itself.
II. If there is a relation from s to s', s different of s' then there is no relation from s' to s.
III. If there is r from s to s' and r' from s' to s'', then there is the relation r''.(r', r) from s to s''.
We denote by s'>>s, if and only if there is a relation r from s to s'.
We call >> the Causal Relation between Physical Systems.
Two more postulates referring to Spacetime Events are necessary:
IV. Spacetime is a non empty set **E**, the elements of which we call events.
V. We say "a physical system s is represented in an event e with the following properties:
a /for any event e there is the physical system s such that s is represented in e.
b /if s is different of s' and they are represented in the same event e then there is no r from s' to s or from s to s'.
c /for any s there is the event e such that s is represented in e.
Now we define the relation between the events, e>e', if and only if the system s being represented in the event e implies the existence of s' represented in e' and it implies that: there is the relation r from s to s', and; if e=e' then s=s'.

We call " > " Causal Relation of Spacetime. We showed that both relations, >> and >, being reflexive, antisymmetric and transitive, are Partial Order Relation.

**Proposition 1**. Causal Relations are Partial Order Relations (POR).

Now the point is in getting a mathematical partial ordered structures (POS), which will represent the causal structure, defined above. So we add another property.

VI. There is a bijection K which maps E on the linear, real, four-dimensional space $R^{1+3}$. So K sets one to one correspondence between the spacetime events of E and the vectors of space $R^{1+3}$, K:E--- $R^{1+3}$ It is the very simplest case.

Let $\mathbf{x} = x^{\wedge}$ (^ = 1, 2, 3)$\in R^3$. and let $x^{\wedge}$ (^ = 0, 1, 2, 3,) $\in R^{1+3}$ be the components in a certain basis. Zero vectors are denoted by **0** and 0 respectively.

We formulate now the causal structural constraints which a map f: $R^{1+3}$ - - R must satisfy, to realize POS that corresponds to the POS defined above for physical systems:

a. for any t$\in$R, t>0, there is x$\in R^{1+3}$ with $x^o = t$, $\mathbf{x} \neq 0$ and f(x) = 0;

b. for any $\mathbf{r}\in \mathbf{R}$, $\mathbf{r} \neq \mathbf{0}$, there are the vectors x and y with $x^o = y^o > 0$, $\mathbf{x} = -\mathbf{y} = \mathbf{r}$ and f(x) = f(y) = 0 f(x) = 0 and $\mathbf{x} = \mathbf{0}$ imply x = 0;

c. for any x,y $\in R^{1+3}$ with $x^o > 0$, $y^o > 0$, f(x) $\geq 0$ and f(y) $\geq = 0$ imply
$$f(x) + f(y) \leq f(x+y),$$
equality is valid if the vectors x and y are linearly dependent,

d'. if f(x) = 0 and f(y) = 0 then f(x+y) = 0 if and only if x and y are linearly dependent,

e. for any x,y$\in$R with $x^o = y^o > 0$, $\mathbf{x} = -\mathbf{y} \neq 0$, f(x) $\geq 0$, f(y) $\geq 0$ imply,
$$f(x) = f(y).$$

The above conditions are independent. However for the convenience of the physical interpretation, we add a condition that follows from the above constraints, lemma 1.[ Consequence of Causality. Preprint DFPD 94 -TH-33 (1994) ]

f. the map f is strictly increasing in the $x^o$ component, in any sub-domain
$$D_y = \{x \mid x \in R^{1+3}, x\, f\, y, \mathbf{x} = \mathbf{y}\}$$
y $\in$R with $y^o > 0$ and f(y) = 0.

Using physical significance of the properties of f we showed that it really represents what we called causal relation between spacetime events.

In other words, we showed that the relation e>e' is equivalent with xfy, where x = k(e) and y = k(e'), and the relation xfy with x $\neq$ y is equivalent to
$$x^0 < y^0$$
$$xfy \Leftrightarrow f(x - y):$$
The xfy expresses in fact the possibility of a relation r between the events s and s' represented respectively in x and y.

**Proposition.2.** The causal maps f form a ring with respect to positive numbers.

**Reference.**

[1] El. Mihul, General concept of Causality. Winter school of Theor. Phys. Karpacz(1970)

[2] E. C. Zeeman "Causality implies the Lorentz group". J. Math. Phys.5 (1964)

[3] El. Mihul. Consequence of Causality. Preprint DFPD 94 -TH-33 (1994)

[4[ C. Gheorghe, EL.Mihul, Commun. Math. Phys.14, 165 (1969),

[5] Fuchs, L: Partially ordered algebraic systems. New York : Pergamon Press 1963.

[6] E. H. Moore, H. L. Smith A general theory of limits. Ammer.J.Math.44 102-121 (1922)

[7] El. Mihul. XII Int. Conf on Selected of Modern Phys. JINR June 2003

[8] Eleonora A. Mihul. UBPub-EPPG/54.May 2005, Lorentz Causality Consequences, Neutrino and Photon, Coulomb and Newton Laws.

[9] E. C. Zeeman The topology of Minkowski Space, Topology Vol.6 161-170 (1967).

[10] Eleonora Mihul UBPub-EPPG/47, April 2000 Lorentz causality and the three neutrinos. Presented at L3 general meeting at CERN April 2000.

[11] C. Gheorghe, EL.Mihul, Commun. Math. Phys. 43, 89-108 (1975). [On the G Geometry of Lorentz Orbit Spaces]

[12] Edvin A. Abbott realizes that the fourth dimension may be real [The Annotated Flat-land: A Romance of Many Dimensions. Edvin A. Abbott Perseus, Cambridge, Mass. 2000.

[13] El .Mihul, Rev. Roum. Phys.28 N1,3, Bucharest (1983).